\documentclass[]{aa}

\usepackage{graphicx}
\usepackage{amssymb}
\usepackage{float}
\usepackage{color}
\usepackage[varg]{txfonts}
\usepackage{epstopdf}
\usepackage{array}
\usepackage{booktabs}
\usepackage[round]{natbib}
\begin{document}

\title {Quasi-oscillatory dynamics observed in ascending phase of the flare on March 6, 2012}

\author{E. Philishvili \inst{1,2}, B.M. Shergelashvili \inst{3,2,5}, T.V. Zaqarashvili \inst{4,2}, V. Kukhianidze \inst{2}, G. Ramishvili \inst{2} , M. Khodachenko \inst{3,6}, S. Poedts \inst{1}, P. De Causmaecker\inst{5}}

\institute{ Centre for Mathematical Plasma Astrophysics, Celestijnenlaan 200B, 3001, Leuven, Belgium\\
 \and
Abastumani Astrophysical Observatory at Ilia State University, Kakutsa Cholokashvili Ave 3/5, Tbilisi, Georgia\\
 \and
 Space Research Institute, Austrian Academy of Sciences, Schmiedlstrasse 6, 8042 Graz, Austria\\
 \and
 Institute of Physics, IGAM, University of Graz, Universit\"atsplatz 5, 8010 Graz, Austria\\
 \and
 Combinatorial Optomization and Decision Support, KU Leuven campus Kortrijk, E. Sabbelaan 53, 8500 Kortrijk, Belgium\\
 \and
Lomonosov Moscow State University, Skobeltsyn Institute of Nuclear Physics (MSU SINP)
Leninskie Gory, 119992, Moscow, Russia\\}

\abstract
{The dynamics of the flaring loops in active region (AR) 11429 are studied. The observed dynamics consist of several evolution stages of the flaring loop system during both the ascending and descending phases of the registered M-class flare. The dynamical properties can also be classified by different types of magnetic reconnection, related plasma ejection and aperiodic flows, quasi-periodic oscillatory motions, and rapid temperature and density changes, among others. The focus of the present paper is on a specific time interval during the ascending (pre-flare) phase.}
{The goal is to understand the quasi-periodic behavior in both space and time of the magnetic loop structures during the considered time interval.}
{We have studied the characteristic location, motion, and periodicity properties of the flaring loops by examining space-time diagrams and intensity variation analysis along the coronal magnetic loops using AIA intensity and HMI magnetogram images (from the Solar Dynamics Observatory(SDO)).}
{We detected bright plasma blobs along the coronal loop during the ascending phase of the solar flare, the intensity variations of which  clearly show  quasi-periodic behavior. We also determined the periods of these oscillations.}
{Two different interpretations are presented for the observed dynamics. Firstly, the oscillations are interpreted as the manifestation of non-fundamental harmonics of longitudinal standing acoustic oscillations driven by the thermodynamically non-equilibrium background (with time variable density and temperature). The second possible interpretation we provide is that the observed bright blobs could be a signature of a strongly twisted coronal loop that is kink unstable.}


\titlerunning{Quasi-oscillatory dynamics}

\authorrunning{Philishvili et al.}

\maketitle

\section{Introduction}

\begin{figure*}
  \includegraphics[width=1.0\linewidth]{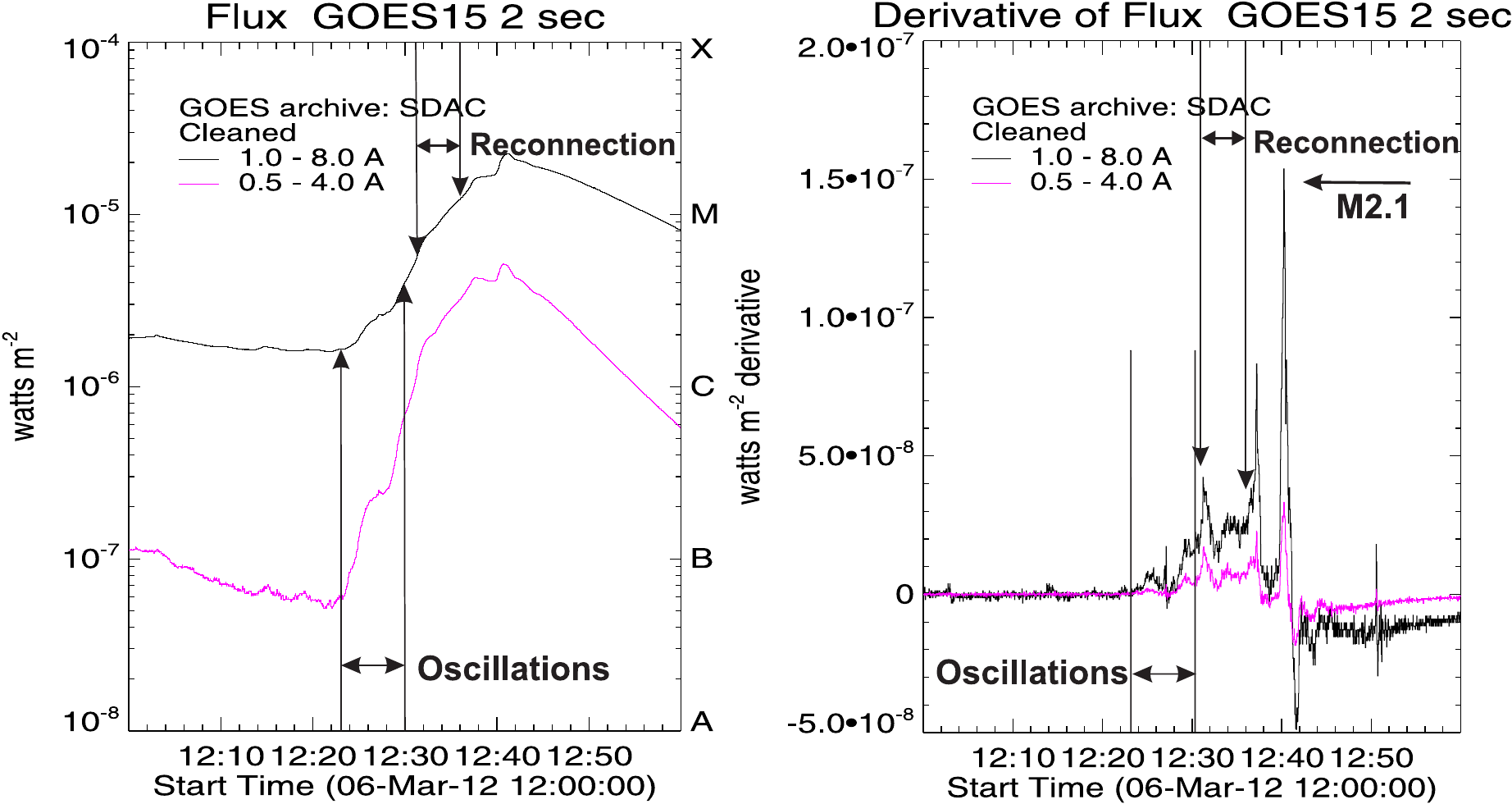}\\  
\caption{GOES soft X-ray flux and flux derivative in the $1-8\;\AA$ and $0.5-4\;\AA$ channels distance-time measurements of the M2.1 class flare. With the arrows we show time intervals of evidence of oscillatory phenomena and magnetic reconnection. The  reconnection starting time corresponds to the peak on flux derivative plot.}
\label{Figure12}
\end{figure*}

\begin{figure*}
  \includegraphics[width=1.0\linewidth]{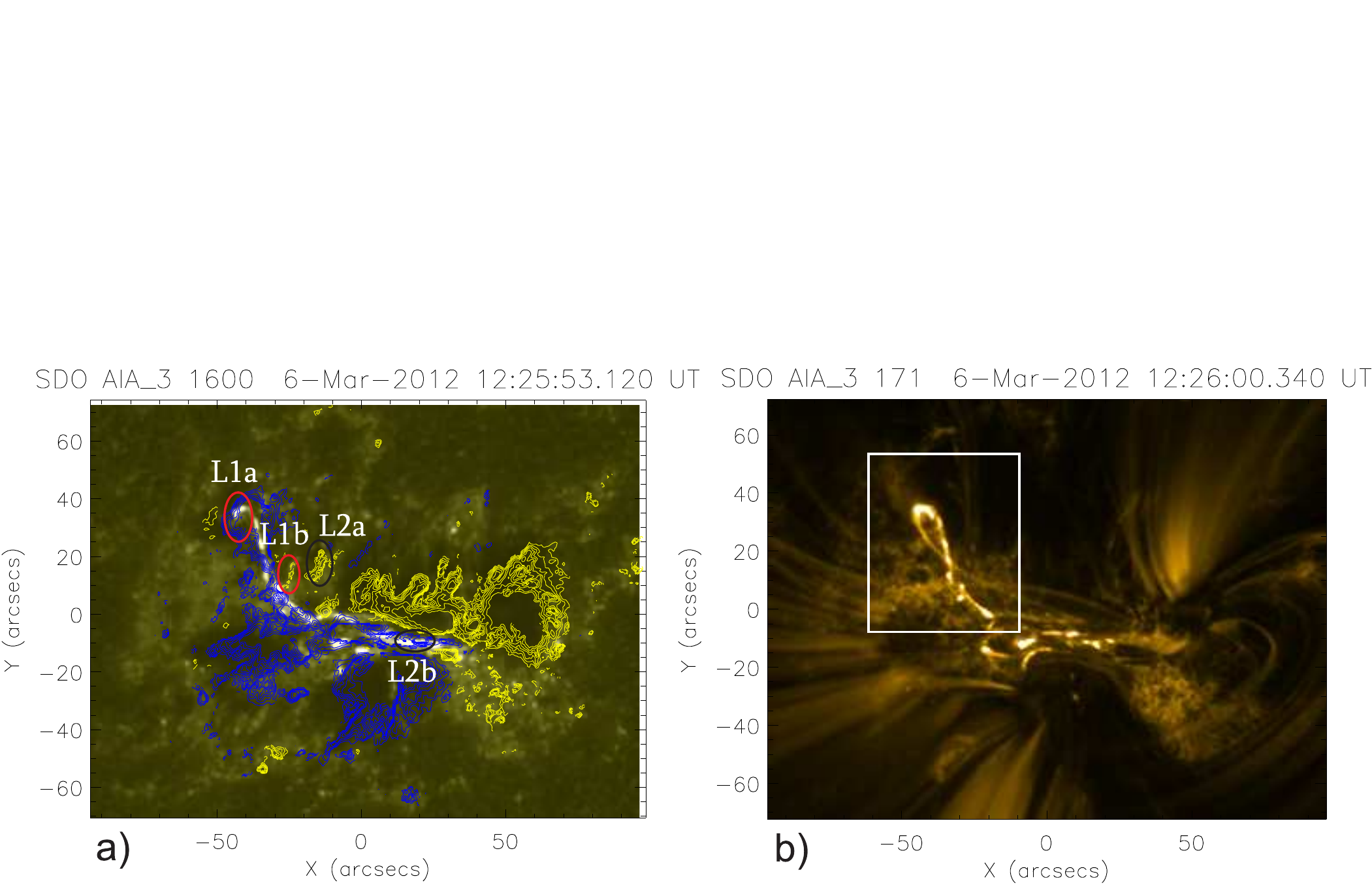}\\
 \caption{Images of AR NOAA 11429 and flaring loops. Panels (a) and (b) show the entire map of the AR in $1600\;\AA$ and $171\;\AA$ emissions, respectively. Blue and yellow areas represent positive and negative values of the LOS magnetic field, respectively. The highlighted magnetic field varies in absolute value between the lower threshold of 500 Gauss and upper limit of 1000 Gauss. L1a and L1b (encircled by red ellipses) show the foot points of the L1 loop; while L2a and L2b (encircled by black ellipses) indicate the foot points of the L2 loop, respectively. In panel (b), the rectangular polygon shows the location of the studied flaring loops.}
  \label{Figure1}
\end{figure*}

\begin{figure*}
\center{\includegraphics[width=1.0\linewidth]{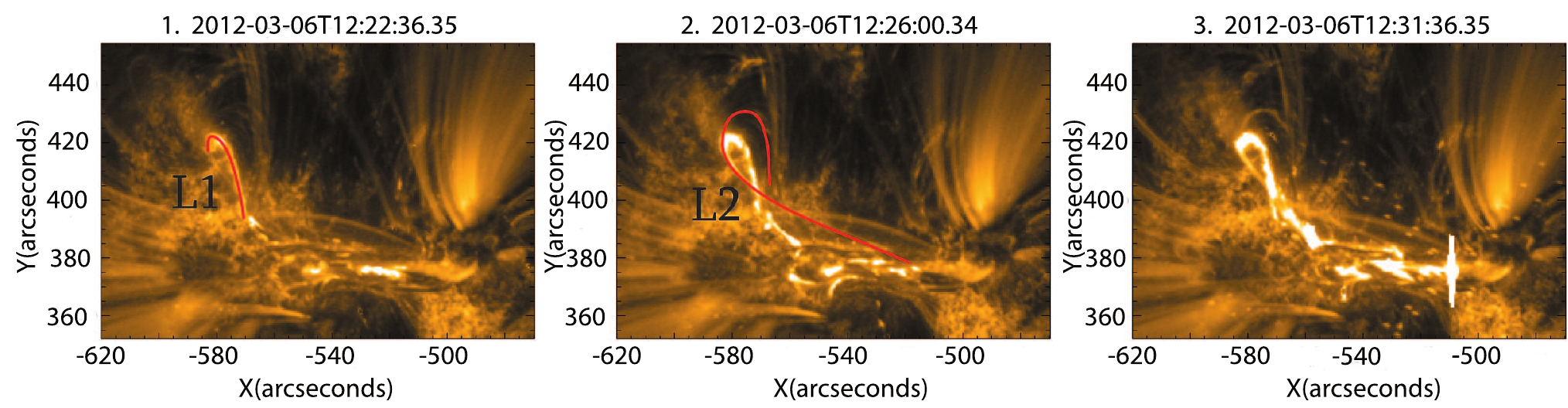}}
\caption{Time sequence of AIA/SDO $171\;\AA$ images of the active region NOAA 11429 (time interval, 12~s; image scale, 0.6\farcs per pixel). The location of the detected flare with respect to\ Heliographic coordinates (in degrees) is: -40, 21. Partial $171\;\AA$ images during 12:22:36 UT-12:31:00 UT  show the coronal loop appearance and the temporal evolution of the flare ascending phase. The red contour in panel 1 represents magnetic loop (L1). Panel 2 shows the L2 loop that appeared, and the last panel demonstrates the time when the L1 and L2 loops widened and the evident magnetic reconnection started between them. }
\label{Figure2}
\end{figure*}

\begin{figure*}
\center{\includegraphics[width=1.0\linewidth,height=5cm]{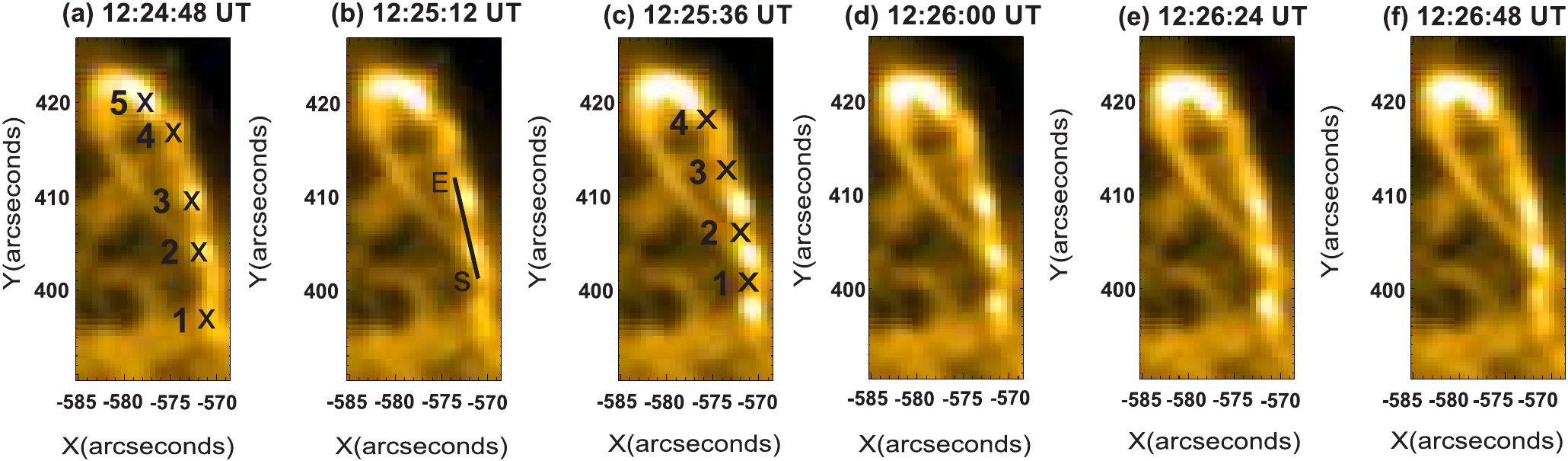}}
\vfill
\center{\includegraphics[width=1.0\linewidth, height=0.2\textheight]{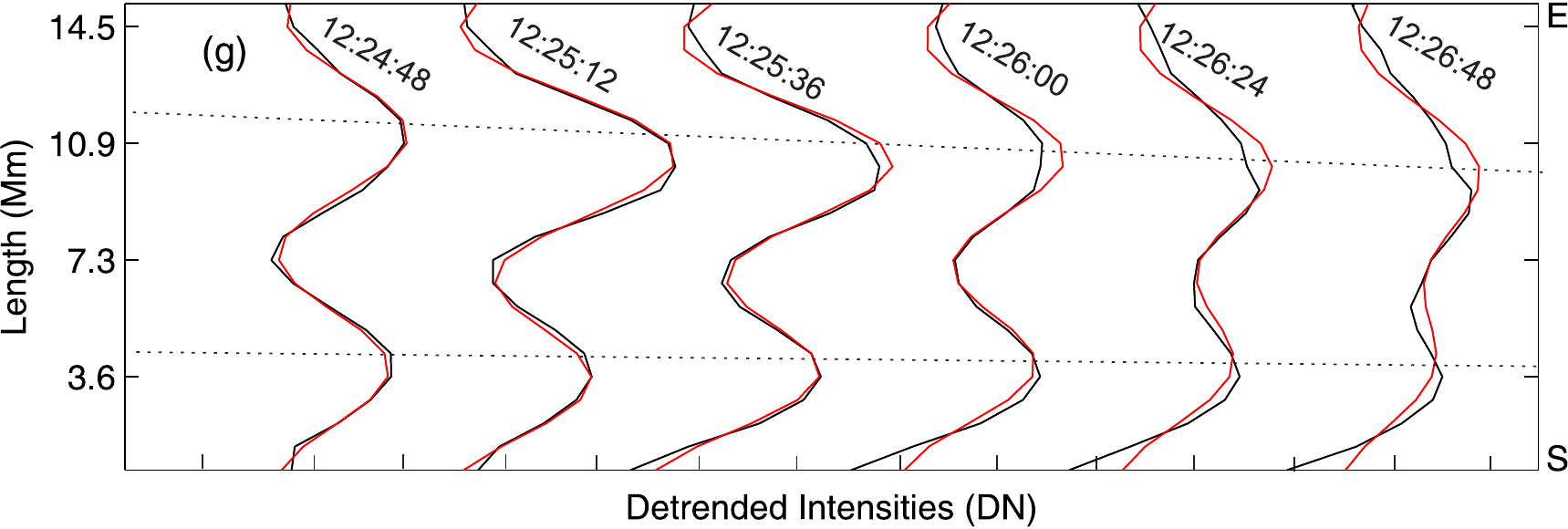}}
\caption{Spatial periodicity of L1 loop and its evolution. Panels (a)-(f) show a temporal sequence of the L1 loop in the period of 12:24:48 UT -12:26:48 UT. Five bright blobs are indicated with crosses on panel~(a). Panel (b) shows a one pixel cut from starting point S to the ending point E, and the corresponding de-trended intensity curves (black) are shown on panel (g), fitted with the function $a\sin(bx+c)$ with the red lines. Intensity varies by $\pm$2000DN. The blobs do not show significant movement. Panel (c) shows four intensity minima points between neighboring bright blobs.}
\label{Figure3}
\end{figure*}

\begin{figure*}
  \includegraphics[width=1.0\linewidth]{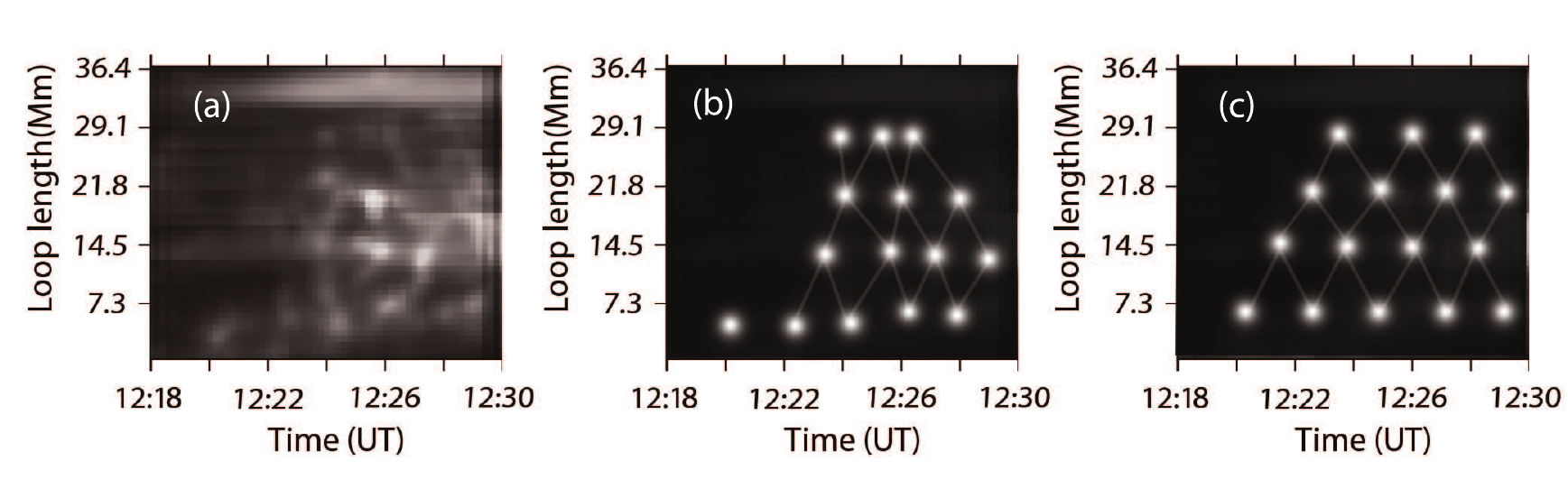}\\
 \caption{Time series for intensity obtained by one pixel slit along the L1 loop using imaging observations in 171$\;\AA$ (SDO/AIA) for  the time period 12:18 -12:30 UT, at 12 sec time intervals. The middle panel  shows schematic diagram of temporal intensity changes. The right panel shows ideal standing mode pattern, that is,\  for oscillations over a background in thermodynamic equilibrium. }
\label{Figure4}
\end{figure*}

\begin{figure*}
  \includegraphics[width=1.0\linewidth]{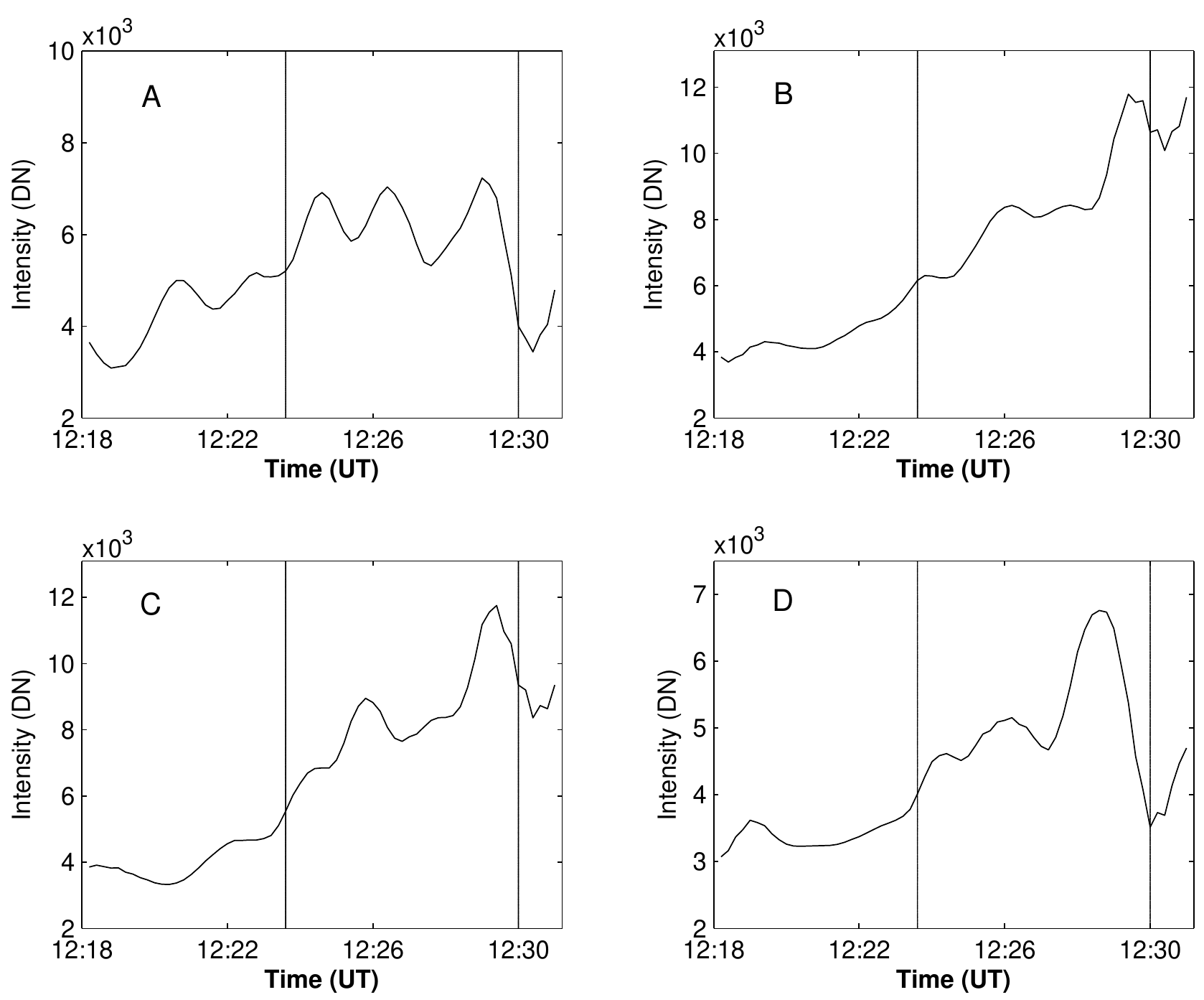}\\
 \caption{Observations of the oscillations within the intensity brightenings. Time series using imaging observations in 171$\;\AA$ (SDO/AIA) for the time period 12:18-12:31 UT, at 12 sec time intervals. Panels A-D show the evolution of intensity for the first four bright blobs (1 to 4), respectively (see Fig.~\ref{Figure3}). The Vertical lines show the time period we selected for the detailed studies of wave characteristics.}
\label{Figure5}
\end{figure*}

\begin{figure*}
\begin{minipage}[h]{1.0\linewidth}
  \includegraphics[width=1.0\linewidth]{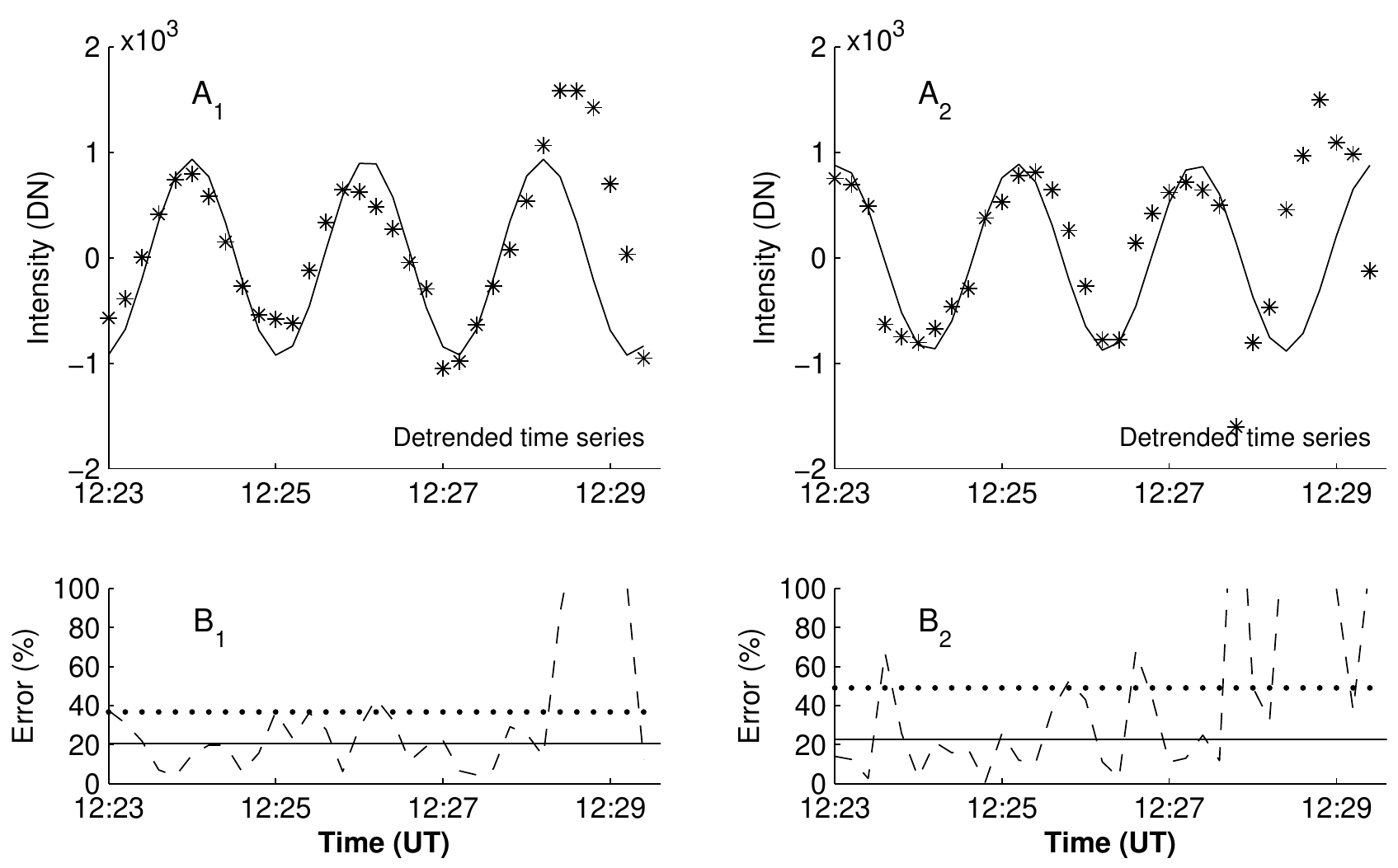}
\vfill
\end{minipage}
\begin{minipage}[h]{1.0\linewidth}
  \includegraphics[width=1.0\linewidth]{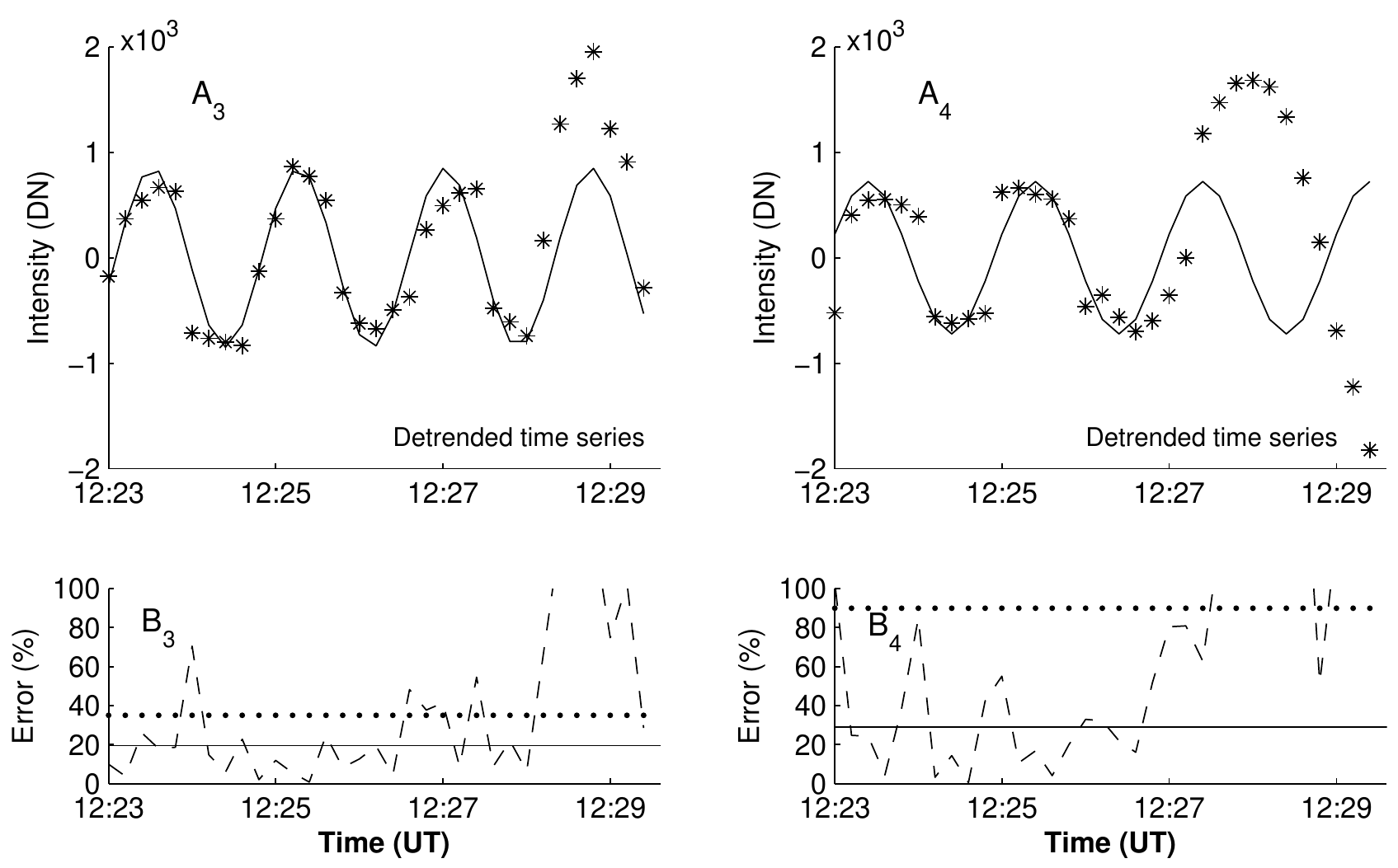}
\vfill
\end{minipage}

 \caption{Analysis of the piecewise de-trended data corresponding to Fig.~\ref{Figure5}.  Panels $A_k, k=1,2,3,4$ show the evolution of de-trended intensity (shown with asterisks) for the first four bright blobs (1 to 4), respectively (see Fig.~\ref{Figure3}) fitted by sin functions (solid lines). Panels $B_k, k=1,2,3,4$ show the fitting errors with dashed lines. The dotted horizontal lines represent the average error for the whole considered time span and the solid horizontal lines show the average error corresponding to the shorter time span up to approximately 12:28 UT.}
\label{Figure6}
\end{figure*}

\begin{figure*}
  \includegraphics[width=1.0\linewidth]{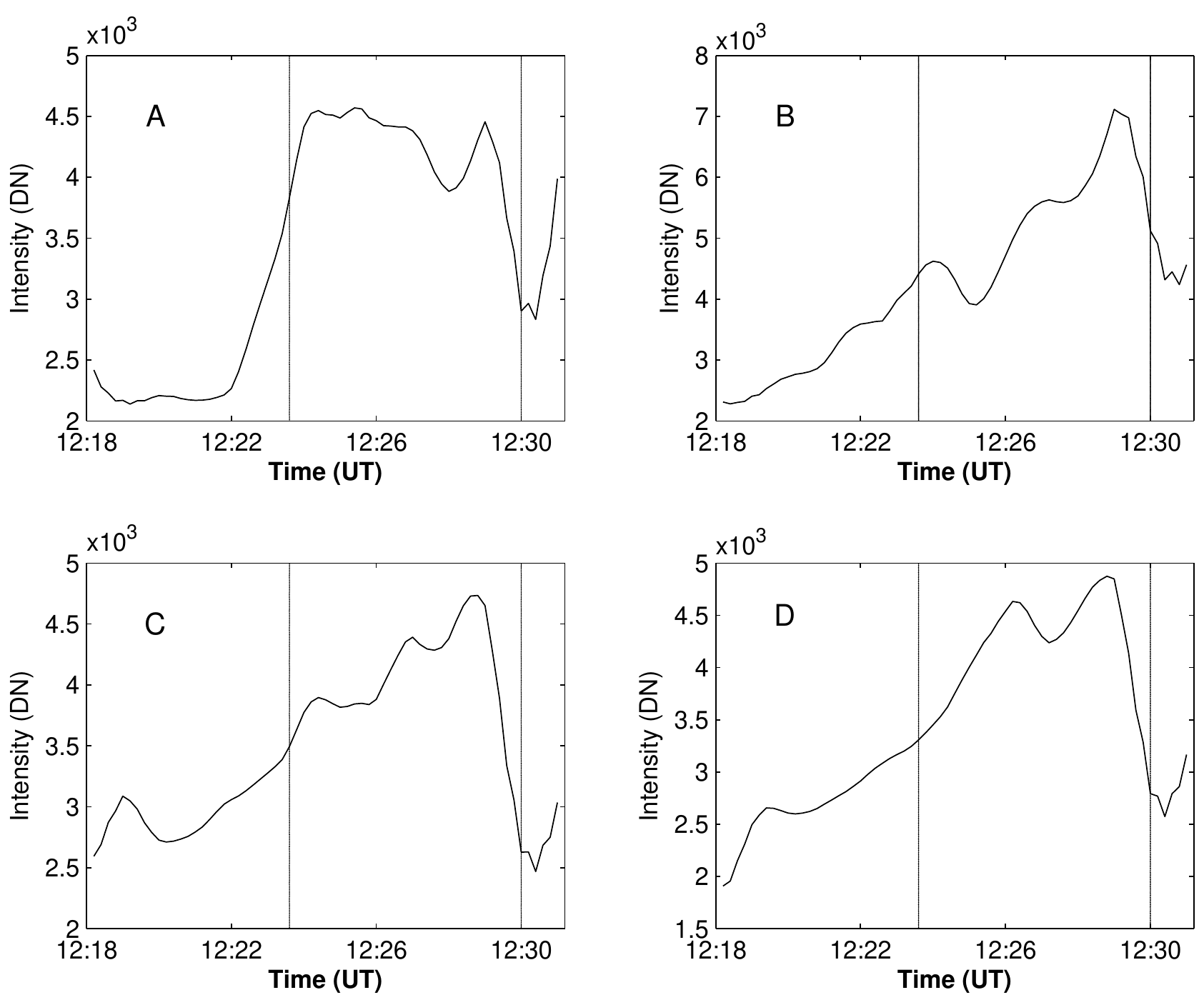}\\
 \caption{Same as Fig.~\ref{Figure5} but for the intensity minima, middle points for neighboring two bright blobs as marked in Fig.~\ref{Figure3} panel (c).}
\label{Figure7}
\end{figure*}

\begin{figure*}
\begin{minipage}[h]{1.0\linewidth}
\includegraphics[width=1.0\linewidth]{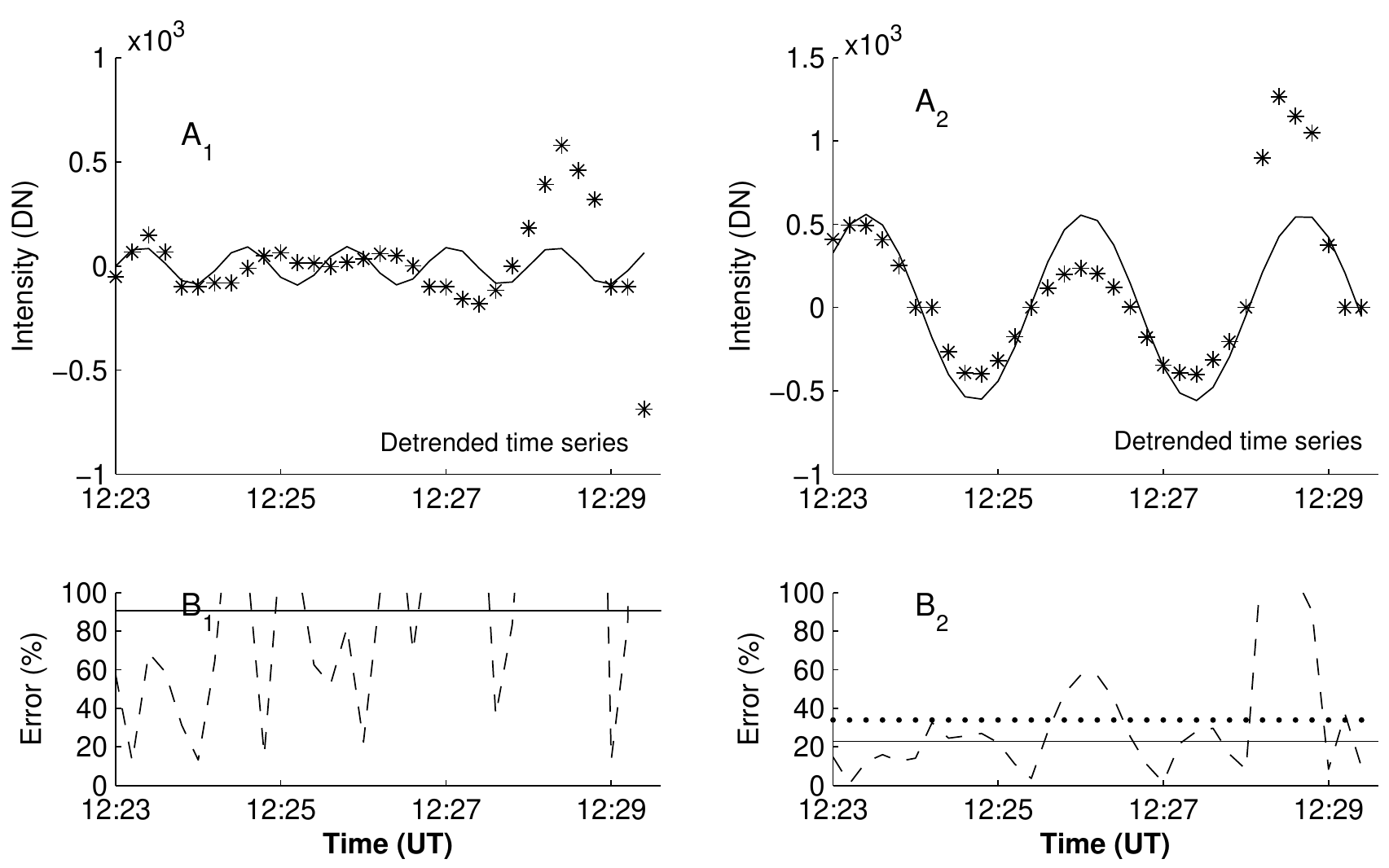}
\vfill
\end{minipage}
\begin{minipage}[h]{1.0\linewidth}
  \includegraphics[width=1.0\linewidth]{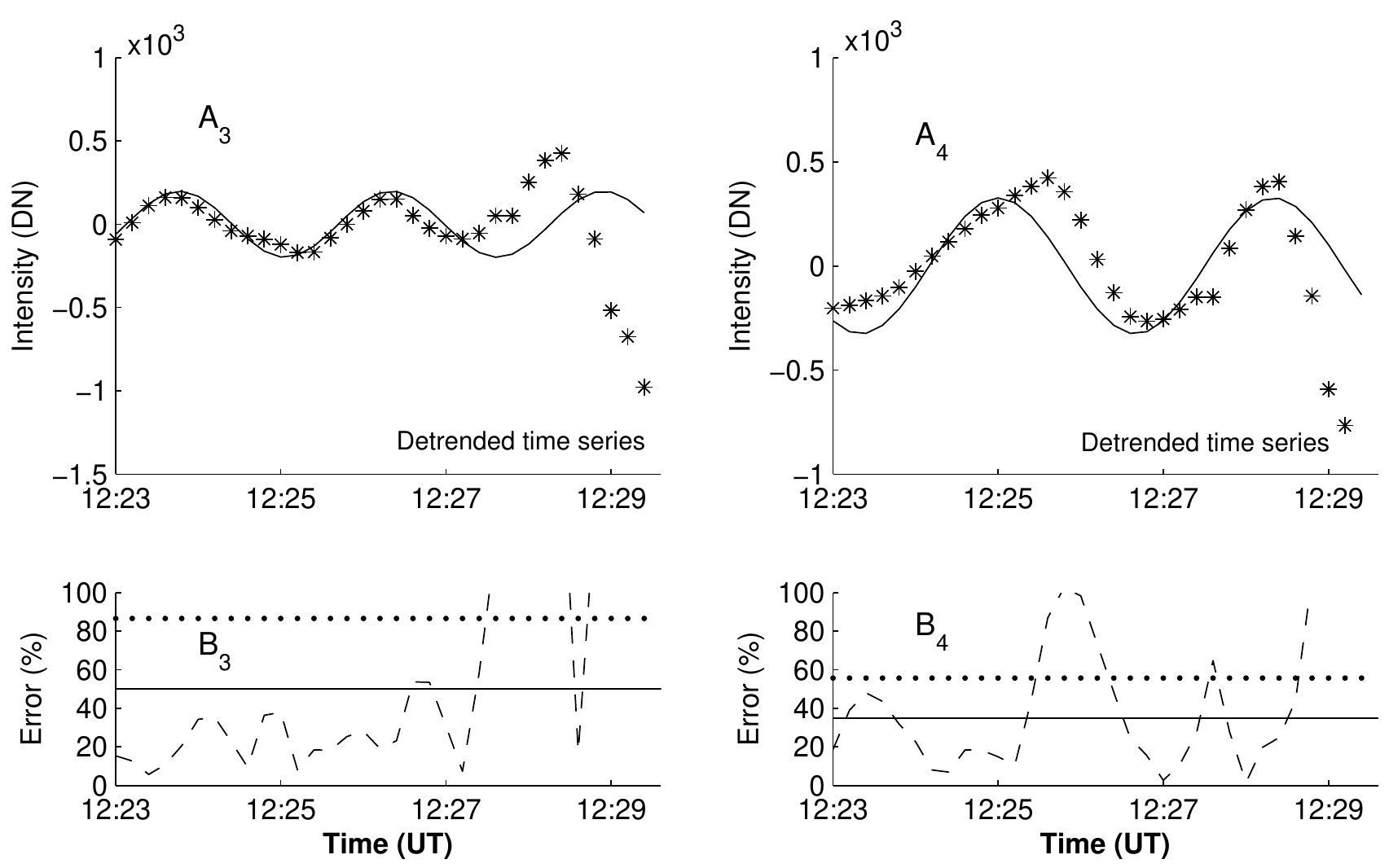}
\vfill
\end{minipage}
 \caption{Same as in Fig.~\ref{Figure7}, but for the intensity minima.}
\label{Figure8}
\end{figure*}

\begin{figure*}
\centerline{%
\includegraphics[width=1.0\textwidth]{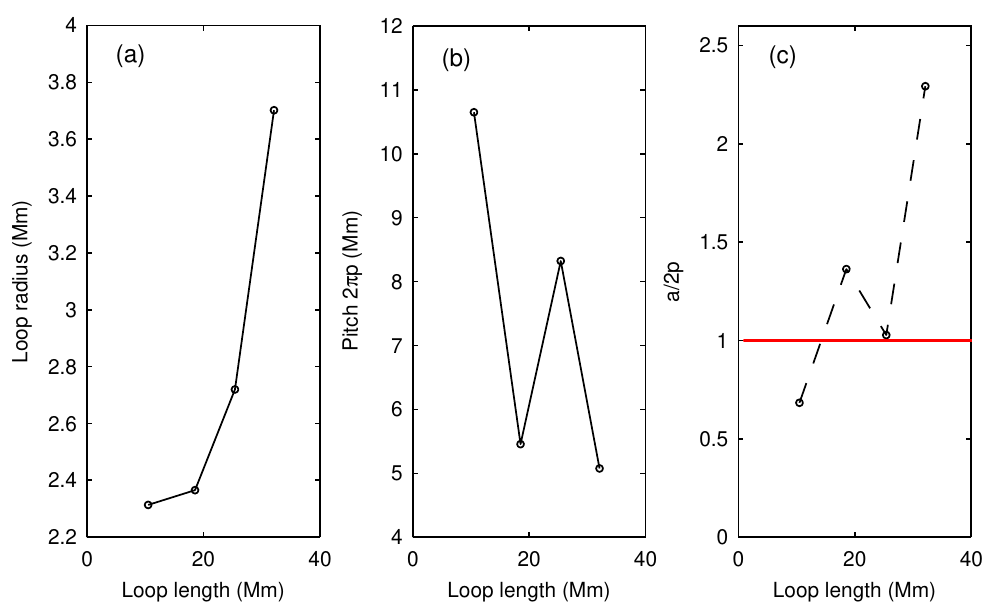}%
}%
\caption{This figure shows the distribution of physical properties along the L1 loop. Panel (a) shows dependence of the loop width on the loop length. Derived radii with the Gaussian Fitting increases from 2.3Mm to 3.7Mm. Panel (b) shows distance between neighboring turns. Panel (c) shows the critical value for kink instability criterion for large aspect ratio loops. Above the critical value of 1, this criterion is satisfied and the occurrence of the kink instability is likely.}
\label{Figure10}
\end{figure*}

The complex magnetic fields in the solar corona represent causes and energy sources for different eruptive and explosive events. In turn, such events are often precursors of large space weather phenomena and play a central role in the formation of local solar weather conditions. The magnetic properties of the ejected plasma clouds strongly affect the space weather related magnetic storms in the magnetosphere of the Earth. A proper observational and analytical study of these phenomena and of the physical mechanisms behind them is important for understanding the solar atmospheric and heliospheric dynamics as part of the unified Sun-Earth system. In fact, this is one of the key tasks of the forthcoming Solar Orbiter (ESA) mission. In the solar and space weather context, flares and other explosive events have been studied by many authors. There is a general concensus that magnetic reconnection plays a key role in the driving mechanism of these events.

Magnetohydrodynamic (MHD) waves and oscillations in various solar structures often lead to instabilities, which in turn can trigger solar flares and coronal mass ejections (CMEs). As a matter of fact, oscillatory regimes of magnetic reconnection can modulate flare processes and show quasi-periodic properties \citep{Kliem2000, Asai2001}.
Quasi-periodic pulsations (QPPs) are associated with flare energy releases in the solar atmosphere, which are observed in radio, soft X-ray (SXR), hard X-ray (HXR), extreme ultraviolet (EUV), and in gamma ray emissions \citep{Roberts1984, Farnik2003, Nakariakov2006, Nakariakov2009, Nakariakov2010}. The period of QPPs ranges from sub-seconds to tens of minutes \citep[see, e.g.,][and references therein]{Tan2008,Nakariakov2009, Karlicky2010, Kupriyanova2010, Su2012}.

Many theoretical models have been proposed to explain the generation of QPPs. The most elaborated model of QPPs considers MHD oscillations, which affect almost all aspects of the flare emission generation. Indeed, QPPs are involved in triggering the magnetic reconnection, modulating the reconnection rate, accelerating and transporting non-thermal electrons, and changing the physical conditions in emitters \citep{Nakariakov2009}. Other models are based on a sandpile system with self-organized critical states \citep{Lu1991,Baiesi2008}, the quasi-stabilized system of non-linear plasmas governed by an oscillatory phase of wave-wave or wave-particle interactions \citep{Aschwanden1987}.
Different MHD oscillation modes have been identified to be responsible for QPPs in a single flaring loop \citep{Nakariakov2003,Melnikov2005,Warmuth2005,Inglis2008,Kupriyanova2010,Kim2012,Kupriyanova2013}.
\citet{Kolotkov2015} studied the QPPs of the microwave emission generated in a X3.2-class solar flare. They found three well-defined intrinsic modes with  mean periods of 15, 45, and $100\;$s. These authors proposed that the $100\;$s and $15\;$s modes are likely to be associated with fundamental kink and sausage modes of the flaring loop, respectively. The $100$s oscillations could also be caused by the fundamental longitudinal mode. The $45$s mode, on the other hand, could be the second standing harmonic of the kink mode.
\citet{Inglis2009} reported a multi-periodic oscillatory event with three distinct periods, namely $28\;$s, $18\;$s, and $12\;$s. They argued that the cause of this multi-periodic event is likely to be a kink mode that periodically triggers magnetic reconnection.
Similar QPPs could be generated by different mechanisms. To discover, understand, and distinguish these mechanisms correctly, detailed and multi-wavelength observations are required. Within the framework of the EU FP7-project SOLSPANET (www.solspanet.eu), we are developing a solar and space-weather knowledge base that will allow such extensive and detailed studies.

In the present paper, we discuss only the ascending phase of the solar flare that was observed by SDO/AIA on March 6, 2012, in NOAA active region 11429. An M2.1-class flare occurred at 12:23 UT, peaked at 12:41 UT and underwent a decay phase until 23:56 UT. From 12:20 UT until 12:36 UT, a variety of phenomena took place, including quasi-periodic oscillations, magnetic reconnection, and rapid plasma ejection that may have triggered the solar flare. Observations of the solar flare and the physical parameters of the flaring loops are described in section~2. Section~3 provides a description of the features of the oscillations and a physical analysis. The discussion and conclusions are given in the last section.

\section{Observations and data processing}
We employ SDO observations to study the dynamics of active region (AR) NOAA 11429. We combined observations from different instruments on board the SDO spacecraft \citep{Pesnell2012} so as to cover the wide range of heights in the solar atmosphere in the vicinity of the flaring region. In particular, the Atmospheric Imaging Assembly (AIA) \citep{Lemen2012} enables continuous imaging of the full disc of the Sun in a temperature range from $\sim 5000\;$K to $\sim 20\;$MK, at 12 sec time intervals and at  spatial resolution of $0.6\;$arcsec per pixel. AIA has two UV and seven narrow EUV passbands; six of the EUV narrowbands are sensitive to different iron ionization lines formed at different temperatures, which are used to study the coronal structure at different heights. The AIA channel in the $304\;\AA$ extreme ultraviolet band shows the first ionization helium (He II) lines, which represents the lower solar atmosphere and is designed to observe areas in the chromosphere and the lower transition region. High-resolution (0.6\farcs) magnetograms were obtained from the Helioseismic and Magnetic Imager (HMI) and these data have been used to observe the photospheric magnetic field of the considered AR. As a matter of fact, HMI measures the full solar disk line of sight (LOS) magnetic fields, vector magnetic fields, and Doppler velocities in the photosphere. We used HMI LOS magnetograms in the Fe~I absorption line with wavelength $6173\;\AA$ at 45 sec time intervals. We also employed observations of the Geostationary Operational Environmental Satellite (GOES) to explore the rate of X-ray radiation related to the studied event. To download, calibrate, and analyze AIA and HMI data, we used standard routines in the Solarsoftware (SSW) package. In order to co-align all  the AIA multi-wavelength images and the HMI magnetograms, we first determined the matching times before rotating the HMI magnetograms to match the fields of view (FOV) of the different instruments.

The considered AR 11429 has released 2 B-, 43 C-, 15 M- and 3 X-class flares (i.e.,\ 61 flares in total) during its
transit over the solar disk. This AR was also the source of several CMEs. Among those 61 flares, only 7 C-, 6 M-, and 3 X-class flares were associated with CMEs. The flare of interest in the present paper was observed near the southwest limb with the UTC-HGS-TOPO (Heliographic Stonyhurst) coordinates: -40, 21. It is important to note that there were no CMEs associated with this particular flare. The soft X-ray flux ($1-8\;\AA$) peak recorded by the X-ray monitor on board GOES reached the value $2.1\times10^{-5}\;$Wm$^{-2}$, the flare magnitude  M2.1. According to the GOES flux plot, the flare had an impulsive rise phase between 12:23 UT and 12:41 UT followed by a gradual decay phase until 12:54 UT (Fig.\ref{Figure12}).

\subsection{Ascending phase of the flare}

Figure~\ref{Figure1} displays the AIA/SDO photospheric image of AR 11429 in $1600\;\AA$ line at 12:25:53 UT, on
March 6, 2012. It is co-aligned with the HMI LOS magnetic field image corresponding to the same moment and the magnetic field contours are overlaid on the image. The blue and yellow colors represent the different polarities of the magnetic field. Ellipses show the position of sunspots where the foot points of the studied coronal loops are probably anchored. Panel~b of Fig.~\ref{Figure1} shows a co-aligned AIA/SDO coronal image in the $171\;\AA$ line at 12:26:00 UT, in which the flaring loops are enveloped by the white rectangle.

General information about the evolution of the flare is shown in Fig.~\ref{Figure2}. This figure shows partial coronal snapshots of AR 11429 observed by the AIA/SDO in $171\;\AA$ line (Fe~ix) between 12:22 and 12:31 UT at different moments. At the moment of the flare start-time, only magnetic loop (L1) was observable, which is shown with the red contour in Fig.~\ref{Figure2} panel~1. After a few minutes, at approximately 12:25 UT, a second flaring loop (L2) brightened (cf.\ red contour in Fig.~\ref{Figure2} panel~2 and in this wavelength, these two loops are hard to separate from one another in the image. The foot points of the L1 loop are associated with magnetic spots L1a (negative polarity) and L1b (positive polarity). The coronal loop L2 is longer and more extended; its left leg is presumably anchored in a small positive magnetic polarity spot L2a and its right leg into the negative polarity spot L2b, as shown in Fig.~\ref{Figure1} panel~(a).

Then, at 12:26 UT, hot plasma was injected along loop L1, which caused its widening. At wavelength $171\;\AA,$  the considered magnetic loop starts to break off asymmetrically at 12:31 UT (panel~3 of Fig.~\ref{Figure2}) and afterwards part of it disappears. Our current analysis covers the interval of time from 12:23 to 12:30 UT.

The evolution of the loops has been observed not only in the $171\;\AA$ line, but also at other wavelengths. Similar behavior is seen in the AIA $193\;\AA$ channel (Fe XII, XXIV) which is designed to observe areas in the corona and hot flare plasma with temperatures $1.2\times10^{6}\;$K and $2\times10^{7}\;$K, respectively.

We observed the same phenomenon at other AIA/SDO wavelengths as well, namely in the  $335\;\AA$ and $211\;\AA$ channels. These channels correspond to the active region corona with temperatures $2\times10^{6}\;$K and $2.5\times10^{7}\;$K, respectively. The magnetic structures at these wavelengths cannot be clearly resolved from the background emission. A blurred configuration can be seen in $94\;\AA$, $131\;\AA$ line images, channels that are used to observe flaring regions with temperatures higher than $6\times10^{6}\;$K.

To estimate the physical parameters of the loop L1, we used different SDO images in the AIA/SDO $171\;\AA$ series. The filters of this channel are designed to provide visibility of high-contrast scenes, such as solar flares, and are especially good for studying faint coronal features.
The most clear image of the loop is seen at 12:25:36 UT. We therefore used this frame to measure the loop length, which amounts to $\sim 36\;$Mm.

We also measured the distance between the foot points of the L1 loop using LOS HMI images $\sim 23\;$Mm$\;= 32\;$pixels. However, taking into account the projection effects and considering the loop as a structure equal or longer than a semicircle, the length of the loop L1 can be estimated as $L>\pi d/2 $ and should be more than 50 pixels (Fig.~\ref{Figure1}).

Therefore, taking into account all these physical parameters and AIA/SDO characteristic temperatures (corresponding mainly to the $171\;\AA$ and $304\;\AA$ wavelengths), we suppose that the observed magnetic loop is located in the lower corona and the temperature of the flaring loop plausibly varies between $6.3\times10^{5}$ and $1.2\times10^{6}\;$K.

 \section{Oscillations}
We detected oscillatory motions in the flaring loop L1. In this section, we first describe the physical parameters of the observed quasi-periodic oscillations and then present two different interpretations of these results in separate subsections.

\subsection{Longitudinal standing acoustic modes}
From the flare onset time 12:23 UT to 12:30 UT in $171\;\AA$ line images, five bright blobs occurred along the loop L1 and these blobs vividly showed periodic brightening behavior over time.

Fig.~\ref{Figure3} displays partial $171\;\AA$ images from 12:24:48 UT to 12:26:48 UT at intervals of 24 sec.
We show these snapshots as illustrative examples for the temporal analysis of the data that was performed for the entire period. In panel~(a) of Fig.~\ref{Figure3}, we see the above mentioned bright blobs, which presumably represent density enhancements within the loop.
We studied de-trended intensity changes along loop L1 using the set of obtained one-pixel cuts. In Fig.~\ref{Figure3}(b),  we show only the line between two neighboring blobs connecting points S and E in panel~(b) of Fig.~\ref{Figure3}. In panel~(g), the corresponding intensities along the slices S-E (black lines) are plotted. The obtained data sets are fitted by a sine function $a \sin(bx+c)$ (red lines) along the same slices. The dotted lines connect the different maxima corresponding to the blobs. The upper blob moved approximately $1.45\;$Mm towards the second blob with velocity v $\sim12\;$km/s, which is negligible within the resolution of current measurements, while the second blob does not show any movement at all. There were also no strong movements detected between the other blobs. Therefore, if we were to interpret these patterns as a kind of body wave motion within the loop, one would definitely assume that we are dealing with standing wave patterns.

The latter assumption needs to be proven by investigating other properties of the observed pattern. For this purpose, the following was carried out: we first smoothed the data with 2 boxcar by the routine `SMOOTH' available in IDL. The SMOOTH function returns a copy of array smoothed with a boxcar average of the specified width and the result has the same type and dimensions as the given array. Then, we made cuts of the loop, locally orthogonal to the loop axes, which are separated by a distance of approximately 1 pixel. Using this method, we obtained 50 cuts  in total between two foot points of the loop (as it is shown in panel~1 of Fig.~\ref{Figure2}). Next, we performed a Gaussian fit of the intensity curves along these cuts. The half width of the Gaussian provided values of the loop local width, which show that the radius of the loop changes between 3 and 5 pixels  along it. Also, the peaks of the Gaussian fits revealed 50 consecutive approximative middle points of the loop enabling us to draw a curved axis of the loop corresponding to a given moment of time, and we recorded the intensity values accordingly at these points. By repeating this procedure for 65 moments of time (snapshots) during the entire period of interest, we obtained a time series of intensity variability at all of the above mentioned  50 middle points of the loop axis.

Further analyzing these data, we next ploted the time-distance diagram as shown in Fig.~\ref{Figure4} panel~(a) from 12:18 UT to 12:30 UT. In Fig.~\ref{Figure5} we display the original time series obtained for only the first four blobs, and in Fig.~\ref{Figure6}, the corresponding de-trended version of the same data is shown, corresponding to the time-distance diagram given in Fig.~\ref{Figure4}. We applied the piecewise linear de-trendisation method. For the fifth blob, that is situated very close to L1a (Fig.~\ref{Figure2}), the oscillatory signal is not clearly resolvable due to the significant tilt of the loop in that part towards the line of sight and the over exposure of the snapshots because of the high local temperature. That is why we did not perform a detailed analysis of this blob, neither within the framework of Fig.~\ref{Figure4} nor in Figs.~\ref{Figure5} or \ref{Figure6}.  However, we still assume that there are five density blobs (marked with crosses in Panel~(a) of Fig.~\ref{Figure3}), and below we make our estimations of modal parameters based on this assumption. Before doing so, we provide a very important description of the mode type recognition method we employed.

From the diagram in panel~(a) of Fig.~\ref{Figure4} it becomes evident that we can clearly resolve at least four consecutive blobs (the fifth one only partially) and the intensity of each of them quasi-periodically oscillating in time. In panel~(b) we show a sketch of the blobs represented by white circles on a black background at the moments when the intensities at the central pixels of the blobs are maximal. It is easy to see that the four rows of circles shown do not oscillate in phase and, moreover, that they are approximately in anti-phase to one another. In panel~(a) there are also clearly resolved whitened ridges of the intensity connecting the bright points in different rows. We represent these by thin vertices connecting the circle nodes of the graph in panel~(b). This observation leads us to the conclusion that the intensity oscillation we observe cannot be due to one fundamental or harmonic mode as this would require the oscillation of the entire loop in phase; and the observed ridges (vertices) indicate density enhancements related to the flow of the plasma between the density antinodes (velocity nodes, since as we observed above, the brightened blobs do not move) of the standing mode pattern. Moreover, as these flows occur along the loop axis, it is very plausible that we see the standing pattern of the longitudinal (acoustic mode), provided that each pair of consecutive blobs oscillates approximately in  phase opposition. Of course, the graph plotted in panel~(b) shows certain deformations of the pattern compared to the classical standing acoustic mode one (as is shown in panel~(c)); and these deformations can be ascribed to the non-standard geometry of the magnetic loop and, more importantly, we are observing these oscillations in the ascending phase of the flare, when drastic rises of temperature and inflows of dense hot plasmas in the loop occur (see Fig.~\ref{Figure5} for each blob separately).

On the one hand, we can assume that the observed high (non-fundamental) harmonic of the standing acoustic mode can be excited and driven by such thermodynamically non-equilibrium source processes. We suppose that the mode can even be excited by multiple non-equilibrium sources simultaneously due to the strong couplings caused by the rapid change in background temperature and density (the analytical description of such couplings, in general, has been given in \citet{Shergelashvili2007}, and for the current particular case a similar modeling is beyond the main scope of this work, which is mostly of the observational kind, and  will thus be addressed elsewhere). On the other hand, the presence of the non-equilibrium background actually causes non-uniform distribution of the plasma characteristics, such as temperature and density, along the loop, resulting in its fragmentation and perhaps leading to noticeable deformations of the standing mode pattern compared to the standard case of oscillations over a background in thermodynamic equilibrium, as sketched in panel~(c). As we can see in Figs.~\ref{Figure5} and \ref{Figure7}, clear coherent oscillatory patterns form at 12:23 UT and last until 12:30 UT (marked with vertical lines). While initially starting from 12:18 UT and continuing until 12:23 UT, the signal is spurious and barely resolvable at all, although the background plasma parameters are already engaged in the uptrend of the started flare (see Fig.~\ref{Figure12}) and the loop is already in non-equilibrium state. One can suppose that this is the very time span when, presumably driven by multiple external sources, high harmonic modes may propagate in different directions to form the standing wave pattern. ~As a matter of fact, two consecutive blobs oscillate in almost
opposite phase, meaning that the distance between the blobs corresponds to half of the wavelength (see detailed calculations below).

In all top panels $A_k, k=1,2,3,4$ of Fig.~\ref{Figure6}, we show the piecewise linearly de-trended time series that enabled us to extract only the well depicted oscillatory part from the original signal. Then, for this analysis, we reduced the total time span to 12:23-12:30 UT,  because after this moment the structure of the observed loop is destroyed as another large reconnection occurs bringing the flare to the next stage of its evolution towards the peak M2.1 class. The analysis of the dynamical processes after 12:30 UT goes beyond the scope of the current study and will be presented in a separate publication. In the present paper, our main focus is simply to understand the properties of the oscillatory phenomenon we observe in the bright blobs. This process starts due to a chain of small reconnection events occurring during the initial stage of flare ascending phase (that started at approximately 12:20 UT) and ending at approximately 12:30 UT. In the mentioned panels $A_k$, the solid lines show the fitted sine function and the de-trended data are presented by asterisks. Figure~\ref{Figure8} is constructed in the same manner, but for velocity antinodes.

Similarly, we plot the original time curves for the intensity minima between neighboring blobs (marked in panel~(c) of Fig.~\ref{Figure3},) in Fig.~\ref{Figure7} separately. The density minima (velocity antinodes) can also be seen to oscillate in phase with the density maxima (velocity nodes). However, the oscillation amplitude of the antinodes is significantly smaller than that of the velocity nodes, which at first glance contradicts the classical theoretical picture of such standing modes. We performed the same analysis for these and Fig.~\ref{Figure8} shows the de-trended time series with the fitted functions.

\begin{table*}[t]
\caption{The parameters of the fitted sinusoidal functions for each node (1-5) and antinode (1-4). Periods, amplitudes, precise error $e_p$ , and overall error $e_0$. The bottom row shows the mean values.}
\label{cxrili}
\centering
\begin{tabular}{c c c c c c c c c}
\hline\hline
& \multicolumn{4}{c}{Velocity nodes} & \multicolumn{4}{c}{Velocity antinodes} \\ \cline{2-9}
& {$P$ (sec)} & {$A$ (DN)} & $e_p$ ($\%$) &  $e_o$ ($\%$)& {$P$ (sec)} & {A(DN)} & $e_p$ ($\%$) & $e_o$($\%$) \\ \hline
1& 126 & 933 & 20 & 37 & 75  & 92  & 90 & 156 \\
2& 128 & 887 & 23 & 49 & 159 & 559 & 22 & 34  \\
3& 105 & 850 & 19 & 35 & 154 & 199 & 50 & 87  \\
4& 120 & 722 & 29 & 90 & 200 & 326 & 35 & 56  \\
5& 103 & 255 & 30 & 115&     &        &    &     \\
\hline
Mean &116 & 729  & 24 & 65 &  147 & 294 & 50& 83\\
\hline
\end{tabular}
\end{table*}

Next, we estimated the significance of the observed oscillation signal with respect to\ the error level. To examine the level of confidence of the observed oscillations, we calculated the standard deviation of the given data (D) from their respective approximated sine function (F) on panels $B_k, k=1,2,3,4$  in Fig.~\ref{Figure6} and Fig.~\ref{Figure8}. We also determined the amplitude for each de-trended curve and calculated the ratio of fitting errors to amplitude for each velocity node and antinode separately (plotted with dashed lines) as follows:

 We evaluated the error between the data (D) and the fitted sine function (F) by using the following formula:

\begin{equation}\label{1}
\centering
e=\frac{\sum^{n}_{i=1}|D_i-F_i|}{A},
\end{equation}

where $n$ is the number of samples and $A$ is the average amplitude calculated from the data for each panel separately. The curves of the temporal variation of the error are represented by dashed lines in the bottom panels $B_k, k=1,2,3,4$. In the same panels, the dotted horizontal line represents the average error for the whole considered time span (overall error $e_0$) and the solid horizontal line shows the average error corresponding to the shorter time span (precise error $e_p$) up to approximately 12:28 UT, where the fitting function manifests the variation of intensity most adequately and, as we see in all top panels after this moment of time, the discrepancy between the fitted and observed data increases noticeably because of changes in the loop configuration itself. The presence of such discrepancy is unavoidable as we fit the sine function with the fixed period and initial phase, while in real data, the period changes significantly due to modification of physical conditions in the loop, such as density and temperature. For instance, we see that in panel $A_1,$ the period of oscillation increases, meaning that locally in this blob, the phase speed (sound speed) decreases. On the other hand, we see in panel $A_2$ that the period decreased after 12:28 UT, indicating that the sound speed increased locally in this blob. The situation is similar in other blobs as well. For this reason, we think that the average error calculated for the time span before approximately 12:28 UT shows the realistic error level of our fittings and the average value of the error for all panels is approximately 24$\%$, while means of the errors for the entire time span show significantly higher percentages (on average 65$\%$), as is expected from the above argumentation. In the case of velocity antinodes, even for the best possible fittings, the error levels for the shorter time spans are on average at least 50$\%,$ with the values for the whole time span reaching up to 83$\%$. These arguments lead us to the conclusion that in the case of blobs, we see real oscillations of the intensity with a significantly high level of confidence. At the same time, in the case of velocity antinodes, the observed average amplitudes are noticeably smaller compared to the case of blobs. Besides, the periods of the oscillations, that we retrieve using the fitting on sine function, show very large discrepancies with one another. Thus, they could not represent a single monochromatic oscillation pattern. Therefore, we assume that the variations of intensity we see in the velocity antinodes are not real oscillations, but instead represent some artefact that could be caused by projection and/or other effects.
The values of the obtained periods, amplitudes, mean errors, and best fitting errors for each curve are given in table~\ref{cxrili}.

The wavelength of the observed quasi-periodic oscillations can be derived as:

\begin{equation}\label{3}
\centering
\lambda=\frac{2L}{N}>\frac{72700}{5}=14 540\;\hbox{\rm km}\approx 14.5\;\hbox{\rm Mm}, \end{equation}

where, $L$ denotes the length of the L1 loop and $N$ is the number of blobs.

Based on the observational facts and the data analysis presented, we suggest that these oscillations with a characteristic period of $116\;$sec could be longitudinal acoustic/slow standing waves. In this case it would correspond to the fifth harmonic, as we detected five bright blobs.

Hence, we can estimate the characteristic phase speed (sound speed) in the loop which should satisfy:
\begin{equation}\label{4}
\centering
C_{ph}\geq\frac{\lambda}{P} \approx125\pm11\;\hbox{\rm km/sec}.
\end{equation}

Furthermore, as the sound speed depends on temperature, using the temperature response of SDO/AIA 171$\;\AA$ and the logic related to the temperature given in section~2, we claim that
\begin{equation}\label{5}
\centering
Cs>\sqrt{\frac{\gamma k_{B} T}{\tilde{\mu} m_{p}}}=120\;\hbox{\rm km/sec},
\end{equation}
where $\mu=0.6$ is the mean molecular mass of hydrogen plasma.
This estimation is in good agreement with our findings. Moreover, if we assume the loop L1 to be a semi-circle. Taking into account the physical properties (the length and the distance between the foot points), it would be located no higher than $25\;$Mm. This indicates that the observed loops are located in the low corona.

\subsection{Kink instability}
In this subsection we present an alternative interpretation of the observed dynamics.
Bright blobs along a coronal loop can also be a signature of twisting of the loop. \citet{Srivastava2010}
showed that the total twist angle related to kink modes can be obtained by:
\begin{equation}\label{1}
\Phi=2\pi N_{twist},
\end{equation}
where, $N_{twist}$ is the number of turns over the tube length. At least five different turns (bright blobs) can be seen along the loop L1. The loop width varies along it. Applying a Gaussian fitting, we derived the loop radius. Further, we measured the
radius at four different points along the L1 loop during the flare. We then calculated the mean values for each point. The radius increases from $\sim2.3\;$Mm to $\sim3.7\;$Mm (Fig.~\ref{Figure10} panel~(a)).

We estimated the mean pitch (2$\pi p$), that is, the distance between
subsequent turns of the magnetic field lines, for each image. The mean pitch decreases from $10.7\;$Mm to $5.1\;$Mm
along the loop, but not gradually (see Fig.~\ref{Figure10} panel~(b)).

As there are five blobs along the LI loop, the total twist angle for the L1 loop is approximately:
\begin{equation}\label{3}
\Phi=10.0\pi,
\end{equation}
in the case of homogeneous distribution.
This means that the L1 loop is strongly twisted and could be vulnerable to kink instability, which eventually may lead to a sudden reconnection with other loops and a consequent M-class flare. In fact, the Kruskal-Shafranov instability criterion yields $\Phi>2.0\pi$. However, for a large aspect ratio loop, that is,\ a loop with a large ratio between the loop length and radius, the critical twist angle increases further \citep{Baty2001}. The approximate aspect ratio for the L1 loop, even
for the narrowest part of the loop, is $L/a\sim 18$ (for the widest part we obtain $L/a\sim 11$), which is relatively large. A normal mode analysis for the large aspect ratio case gives the instability criterion  $a > 2p$ \citep{Dungey1954, Bennett1999}, which also yields a large twist angle. Figure\ref{Figure10}(c) shows that this criterion is only not satisfied for the narrowest part of the loop, close to the L1b foot point. Increasing the radius along the loop increases the possibility for kink instability to occur, with the consequent loop interruption.
If we assume that the radius and pitch do not change along the loop and we take mean values of the radius and pitch, the $a/2p >1$ instability criterion is also satisfied for a symmetric cylinder.

\section{Conclusions}
We studied the well-resolved single loop shape L1 using the imaging observations of SDO/AIA and SDO/HMI(6173$\;\AA$) and estimated its length to be $\sim 36\;$Mm, while the loop half-width increases from $\sim 2.3\;$Mm to $\sim 3.7\;$Mm. From the flare starting time, before the reconnection between the L1 and L2 loops occurs, we observed five bright blobs in the loop that show oscillatory patterns in intensity. We analyzed time-distance diagrams along the loop and showed that the blobs do not move. Further, we performed an intensity variability analysis for each of the bright blobs and also for the dark points in between them. We estime the corresponding quasi-periodic oscillation periods to range from 103 sec to 128 sec for the blobs. We derived the mean value to be 116 sec. We found that the bright blobs oscillate in anti-phase with each of their neighbors and, moreover, we observed the whitened ridges of intensity connecting the bright blobs.

According to our findings, we interpret this event as follows with two different possible scenarios:

We intuitively ascribe the observed oscillations to longitudinal standing acoustic waves. If we take into account the  physical parameters of the loop and the distribution of the bright blobs along it, we can assume that we see a high harmonic, specifically the fifth harmonic, of the standing longitudinal acoustic mode. We suppose that this high mode can be excited by multiple non-equilibrium sources due to the strong couplings related to the rapid change in background temperature and density. Further, we suggest that the characteristic phase speed that we estimated is in accordance with the sound speed in the low corona.

Alternatively, we suggest that the bright blobs could also be the signature of a strongly twisted magnetic loop that is vulnerable to kink instability. We calculated the total twist angle for the L1 loop for a homogenous distribution and found $\Phi=10.0\pi$, which is much larger than the Kruskal-Shafranov instability criterion that gives $\Phi>2.0\pi$. The aspect ratio for the widest part of the loop is $L/a \sim 11$ and for such a large aspect ratio, the instability criterion is given by $a > 2p$. If we consider the L1 loop to be a symmetric cylinder, this criterion is easily satisfied. We derived the mean pitch (the distance between subsequent magnetic field line turns) and their corresponding loop radii for different parts along the loop. As a result, this criterion is satisfied with the increase of the radius, which enhances the kink instability leading to magnetic reconnection and consequent energy release. This in turn may have led to the observed M2.1 class flare.

We also observed a decaying flow with a velocity in the range $60-180\;$km/sec. This flow seemingly indicates the plasma jets related to the reconnection of the magnetic field lines. These flows might play the role of key indicators of the reconnection, and the related physical background of these processes will be studied in a separate investigation.

\begin{acknowledgements}
The work has been realized within the framework of Shota Rustaveli National Science Foundation grant \# DI-2016-52. The work was also supported by Shota Rustaveli National Science Foundation PhD student's research (individual) grant \# DO/248/6-310/13 and grant for young scientist for scientific research internships abroad \# IG/46/1/16. The work was supported by European FP7-PEOPLE-2010-IRSES-269299 project - SOLSPANET. Work of B.M.S. and M.L.K. was supported by the Austrian Fonds zur Fïrderung der wissenschaftlichen Forschung under project \# FWF25640-27 and Leverhulme Trust grant IN-2014-016. M.L.K. additionally acknowledges the support of the FWF projects I2939-N27 and P25587-N27 and the grants No.16-52-14006, No.14-29-06036 of the Russian Fund for Basic Research. The work of T.Z. was supported by the Austrian Fonds zur Forderung der wissenschaftlichen Forschung under project P26181-N27.
\end{acknowledgements}

\bibliography{mybibelene}
\end{document}